\documentclass[
    ,final            
  ,sort&compress
  ]
  {aipproc}
\usepackage{amsmath}
\usepackage{units}
\usepackage{graphicx}
\usepackage{braket}
\usepackage{xfrac}
\usepackage[normalem]{ulem}
\usepackage{booktabs}
\usepackage{tabularx}
\newcommand{\op}[1]{\ensuremath{\hat{#1}}}

\newcommand{\x}{\displaystyle x}
\newcommand{\y}{\displaystyle y}
\newcommand{\z}{\displaystyle z}
\renewcommand{\x}{x}
\renewcommand{\y}{y}
\renewcommand{\z}{z}

\newcolumntype{Y}{>{\centering\arraybackslash}X}

\layoutstyle{8x11double}
\begin{document}
\setlength{\AIPhlinesep}{0pt}
\setlength{\abovedisplayskip}{6pt}
\setlength{\belowdisplayskip}{6pt}
\setlength{\abovedisplayshortskip}{0pt}
\setlength{\belowdisplayshortskip}{4pt}

\addtolength{\textfloatsep}{-18pt}
\addtolength{\dbltextfloatsep}{-12pt}

\title{Development and Evaluation of a Quantum Interactive Learning Tutorial on Larmor Precession Of Spin}

\classification{01.40Fk,01.40.gb,01.40G-,1.30.Rr}
\keywords      {quantum mechanics, time-dependence of expectation values, Larmor Precession}

\author{Benjamin R. Brown}{
  address={Department of Physics and Astronomy, University of Pittsburgh, Pittsburgh, PA 15260}
}

\author{Chandralekha Singh}{
}
\date{12 May 2014}

\begin{abstract}
We conducted research on student difficulties and developed and evaluated a quantum interactive learning tutorial (QuILT) on Larmor precession of spin to help students learn about time-dependence of expectation values in quantum mechanics. The QuILT builds on students' prior knowledge and strives to help them develop a good knowledge structure of relevant concepts. It adapts visualization tools to help students develop intuition about these topics and focuses on helping them integrate qualitative and quantitative understanding. Here, we summarize the development and preliminary evaluation.
\end{abstract}

\maketitle
\section{Introduction}
Quantum mechanics (QM) is a particularly challenging subject for undergraduate students. Based upon the research studies that have identified difficulties~\cite{singh2001,zollman2002,wittman2002,singh2008,zhu3,ajmason,marshman} and findings of cognitive research, we have developed a set of research-based learning tools to help students develop a good grasp of QM. These learning tools include the Quantum Interactive Learning Tutorials (QuILTs)~\cite{ptoday2006,singh20082,zhu1,zhu2}. 

Here, we discuss the development and evaluation of a QuILT employing Larmor precession of spin as the physical system to help upper-level undergraduate students in QM courses learn about time-dependence of expectation values. Through a guided approach to learning, the QuILT helps students learn about these issues using a simple two state system. Generally, the expectation value of an observable Q evolves in time because the state of the system evolves in time in the Schr\"odinger formalism.  
	If an operator \op{Q} corresponding to an observable $Q$ has no explicit time dependence (assumed throughout), taking the time derivative of the states in the expectation value and making use of the Time-Dependent Schr\"odinger's Equation where appropriate yields Ehrenfest's theorem:
	\begin{equation}
	\frac{\mathrm{d}}{\mathrm{dt}} \braket{Q(t)} = \braket{\psi(t) | [\op{Q},\op{H}] | \psi(t)} / {i \hbar} \label{ehrenfest}
	\end{equation}

	Working on the QuILT, students learn that a particle with a magnetic moment may exhibit Larmor precession if an external magnetic field is applied. 
For pedagogical purposes, the simplest case of a spin \nicefrac{1}{2} system, which is a two-state quantum system, is chosen in the QuILT to help students learn this challenging topic.  In this case, spin operators in any basis are $2\times2$ matrices, which are least likely to cause cognitive overload to students.  Despite the simplicity of the system, students learn to reason that the knowledge and skills they acquire in this context apply to a broad range of quantum systems.
	
\section{Exploring Student Difficulties}
Before the development of the QuILT, we investigated student difficulties with time dependence of expectation values using open-ended surveys and multiple-choice questions in order to address them in the 
QuILT~\cite{marshman}. Here we summarize the findings from a written survey given to 39 students:

	\textbf{Difficulty with the relevance of the commutator of the operator corresponding to an observable and the Hamiltonian: \label{H}}
	A consequence of Ehrenfest's Theorem is that if an operator \op{Q} corresponding to an observable Q commutes with the Hamiltonian, the time derivative of $\braket{Q}$ is zero, regardless of the state.  However, 46\% of students did not realize that since the Hamiltonian governs the time-evolution of the system, any operator \op{Q} that commutes with it must be a constant of motion and its expectation value must be time-independent.

	\textbf{Difficulty recognizing the special properties of stationary states:} If the magnetic field is along the z-axis, all expectation values are time independent if the initial state is an eigenstate of \op{S_z} because it is a stationary state.   However, 51\% of students surveyed incorrectly stated that $\braket{S_x}$ and $\braket{S_y}$ depend on time in this case.  One common difficulty includes reasoning such as ``since the system is not in an eigenstate of \op{S_x}, the associated expectation value must be time dependent'' even in a stationary state.  Another very common difficulty is reasoning such as ``since \op{S_x} does not commute with \op{H}, its expectation value must depend on time'' even in a stationary state.
		
	\textbf{Difficulty distinguishing between stationary states and eigenstates of operators corresponding to observables other than energy:}
	Any operator corresponding to an observable has an associated set of eigenstates, but only eigenstates of the Hamiltonian are stationary states because the Hamiltonian plays a central role in time-evolution of the state.  However, many students were unable to differentiate between these concepts. For example, for Larmor precession with the magnetic field in the z direction, 49\% of students claimed that if a system is initially in an eigenstate of \op{S_x} or \op{S_y}, the system will remain in that eigenstate.  A related common difficulty is exemplified by the following comment from a student: ``if a system is initially in an eigenstate of \op{S_x}, then only the expectation value of $S_x$ will not depend on time.''  
	
\section{Development of the QuILT}
	Based upon a theoretical task analysis of the underlying knowledge from an expert perspective and common student difficulties found via research, a preliminary version of the QuILT with a sequence of guided questions along with pre-/posttests was developed. The pretest is to be administered after traditional instruction on Larmor precession and the posttest after students work on the QuILT. The QuILT strives to help students build on their prior knowledge and develop a robust knowledge structure related to the time-dependence of expectation values. Students have to actively think through the answers to each question in the guided approach to learning and the latter questions build on the preceding questions.  
	Some questions ask students to justify whether statements that incorporate common student difficulties are correct or not. The QuILT adapts computer simulations~\cite{osp} in which students can manipulate the Larmor precession setup to predict and observe what happens to time-dependence of expectation values for different observables, initial quantum states, and orientations of the magnetic field. In particular, students have to make predictions about whether the expectation values depend on time in various situations and then they engage via simulations of Larmor precession in learning whether their predictions are consistent with the observation.  Then, students are provided with guidance and scaffolding support to help them reconcile the differences between their predictions and observations and to help them assimilate and accommodate relevant concepts. The components of the simulations students use in the QuILT to check their predictions and reconcile the differences include tools to orient the magnetic field in a particular direction, tools for state preparation, quantum measurement, and particle detection.  
	
	Different versions of the QuILT were iterated with five physics faculty members (experts) several times to ensure that they agreed with the content and wording of the questions. We also administered it individually to fifteen graduate students and upper-level undergraduate students in semi-structured think-aloud interviews to validate the QuILT for clarity, student interpretation, appropriateness of the guided sequence and in order to address any emergent issues. Students were first asked to think aloud as they answered the questions to the best of their ability without being disturbed. Later, we probed them further and asked them for clarification of points they had not made clear. Since both undergraduate and graduate students exhibited the same difficulties, we will not differentiate between the two groups further. Overall, the interviews focused on ensuring that the guided approach was effective and the questions were unambiguously interpreted. Modifications were made based upon each feedback. 

\section{Results from the Pre-/Posttests}
	Once we determined that the QuILT was effective in individual administration, it was administered in class. Over several years, from 2008 to 2013, students in four junior-senior level quantum mechanics courses, totaling 72 students, were first given the pre-test after traditional instruction.  They then worked through the QuILT in class and were asked to complete whatever they could not finish in class as homework.  Then, the post-test was administered in the following class which has analogous questions to the pre-test (except, e.g., ${S_x}$ and ${S_y}$ are swapped in various questions).  One class with 18 students (included in the 72 students) was also given four matched questions on their final exam to investigate retention of these concepts more than two months later.
	
	To assess student learning, both the pre-test and post-test were scored based on two rubrics developed by the two researchers jointly.  The ``Strict" rubric only gives credit if the student provides a completely correct answer.  A ``completely correct'' answer is defined as an answer in which both the correct answer is given and it is supported with correct reasoning.  Correct reasoning can be either quantum mechanical or classical in nature (discussed later). Good performance on  this  strict scoring is likely to reflect an expert-like performance.
	
	A second rubric, referred to as ``Partial", was developed to give partial credit in scoring students' pre-test and post-test performances.  This rubric gives half credit for a correct answer (regardless of the reasoning) and half credit for correct reasoning (again, either correct quantum mechanical or classical reasoning is considered valid).  This rubric is consistent with the common traditional methods of grading student performance.
More than 20\% of the sample was scored individually for inter-rater reliability  and the scores of the two raters agreed to better than 90\%.  For assessing retention, the same rubric was used to score matched questions on the final exam which was administered more than two months later.
	

\begin{table}
\caption{The average pre-/posttest scores of 72 students on each of the six items.  For item 6, students were asked if there was precession in a given case and asked for an example in which precession does not take place. For this question (item 6), there is no ``Partial'' grading scheme.}
\label{averageImprovementDetailed}
\begin{tabularx}{\columnwidth-4pt}{@{}|l | c  Y  | c  Y |@{}}
				\cline{2-3}					\cline{4-5}
\multicolumn{1}{l|}{}			&	\multicolumn{2}{c|}{Strict}	&	\multicolumn{2}{c|}{Partial}\\
\hline
    Item 		&Pre	(\%)	&Post (\%)	&Pre (\%)		&Post (\%)	\\
    1			&60.6	&80.6		&63.6		&81.3		\\
    2			&57.6	&83.3		&66.7		&86.8		\\
    3			&50.0	&72.2		&56.1		&79.9		\\
    4			&45.5	&72.2		&52.3		&77.8		\\
    5			&27.3	&55.6		&34.8		&59.7		\\
    6			&50.0	&70.8		&{*}			&{*}			\\
\hline
Combined		&48.4	&72.5		&54.2		&76.3		\\
\hline
\end{tabularx}
\end{table}

	Table \ref{averageImprovementDetailed} shows the average scores on each question on the pre-test and post-test using both the strict grading scheme and partial grading scheme.  The average pre-/post-test performances were compared using a t-test; all improvements (using both the strict and partial rubrics for each question) were statistically significant.   Table \ref{averageImprovementDetailed} shows that the pretest scores were lower under the strict grading scheme than under the partial grading scheme. 

Items 1, 2, 3, and 5 on the pre-/post-tests all ask questions of the following form for a given system: Given a specific initial state, does the expectation value of a certain observable $Q$ change with time? Students were asked to explain their reasoning in each case. These questions are equivalent to asking whether or not the right hand side of Eq. \eqref{ehrenfest} is zero in that case. Items 3 and 5 are specifically designed to probe the knowledge of the two distinct general conditions when the expectation value is time-independent (when $\hat Q$ commutes with the Hamiltonian \op{H} or when the initial state is a stationary state). On Item 6, students were given a system initially in an eigenstate of $\op{S_{\x}}$ and asked whether precession occurs in this case (it does) and why, which prompts students to connect classical and quantum mechanical explanations. 
Below, we discuss how student difficulties were reduced after working on the QuILT.

	\textbf{Difficulty with constants of motion and their significance:} On Item 3,  $S_{\z}$ is a constant of motion because \op{S_{\z}} and  \op{H} commute for the given system.  Therefore, $\Braket{S_{\z}}$ is time-independent regardless of the state. Under the strict rubric, the average score on this question improved from 50\% to 72\% from pre-test to post-test.

	\textbf{Difficulty with properties of a stationary state:}  On Item 5, under the strict rubric, average scores improved by $28\%$ from pre-test to post-test.  However, even in the post-test, some students struggled with the fact that in a stationary state, the expectation values of all observables are time-independent.

	\textbf{Difficulty distinguishing between stationary states and other eigenstates:} On Items 1, 2, 4, and 6, student performance under the strict rubric improved by 25\% on average.
	In Item 4, students consider a system which is initially in a non-stationary eigenstate, e.g., of $\op{S}_{\x}$ (not an eigenstate of \op{H}), and they are asked if the state remains an eigenstate of $\op{S}_{\x}$.  With the strict rubric, the score on this question improved from 45\% on the pre-test to 72\% on the post-test.
	Items 1 and 2 ask students to consider a system initially in an eigenstate of $\op{S}_{\x}$, and ask whether $\braket{S_x}$ and $\braket{S_y}$ depend on time when the magnetic field is in the z direction. Since the initial state is not a stationary state and neither operator commutes with \op{H}, both expectation values are time-dependent.
	While average scores and improvements are approximately equal for these two items (see Table \ref{averageImprovementDetailed}), the common reasoning provided for incorrect answers was different.  On Item 1, in which the initial state is an eigenstate of the operator $\op{S}_{\x}$, 19 out of 36 students who claimed that the expectation value is "time independent" incorrectly stated that the system was in a stationary state.  However, on Item 2 involving time dependence of $\braket{S_{\y}}$ when the initial state is an eigenstate of $\op{S}_{\x}$, very few (3 of 25) students stated that the system was in a stationary state.  Instead, most of the students who answered Item 2 incorrectly  (18 of 25) provided reasoning based on non-commutation of \op{S_{\x}} and \op{S_{\y}}.  For example, one student stated  ``No.  Because it is initially in an eigenstate of $\op{S}_{\x}$.  $\op{S}_{\x}$, $\op{S}_{\y}$ are mutually incompatible observables.''  Synthesis of incorrect responses on Items 1 and 2 suggests that students incorrectly argued that a stationary state is not a property of the system, but that quantum states are stationary \emph{with respect to} a particular observable in question. Also, under the strict rubric, scores on Item 6, designed to probe overall understanding of Larmor precession improved from 50\% to 71\%.
	
In addition to pre-/post-tests, four matched questions (same as four questions on the pre-test/post-test) were administered as part of the final exam to 18 students in a particular year (subset of the 72 students).  Under the strict rubric, the average cumulative scores for this group on matched questions increased from 32\% on the pre-test to 82\% on the post-test, and was 85\% on the final exam.

	\subsection{Quantum vs. Classical Reasoning}
	Most questions posed to students on pre-/post-tests about the time-dependence of expectation values in the context of Larmor precession allowed either a quantum mechanical or classical explanation.  This situation presents an opportunity to investigate and compare the rate at which students adopt and correctly use the new ``Quantum'' framework learned, versus relying primarily on ``Classical'' reasoning.   Explanations considered to be quantum reasoning include statements about operators, their commutation relations, states and their time evolution via the Schr\"odinger equation.  Statements based on whether or not a system will exhibit precession were counted as classical reasoning. Although discussions with some physics faculty suggested that they value quantum reasoning more in the context of a quantum system, appropriate classical reasoning was considered correct in this investigation. 
	
	  An example of a response that would be categorized as quantum reasoning is: ``Yes, as by [Ehrenfest's] theorem.  Since \op{S_{\x}} does not commute with \op{H} and since [it] is not a stationary state, the derivative of the expectation value of $S_{\x}$ with respect to time is not zero.'' An example of classical reasoning when asked if $\Braket{S_{\z}}$ depends on time is: ``Yes, because the magnetic field causes an angular frequency $\omega$ and precession about $z$ axis.'' An example of a response which exhibits both classical and quantum reasoning when asked if an expectation value is time dependent is: ``No, since there is no longer precession about the $z$-axis.  Eigenstates of \op{S_{\z}} are stationary states.'' These types of reasoning were included in both quantum and classical categories.

	 A rubric was developed to categorize student statements pertaining to their reasoning as either ``Quantum'' or ``Classical'' in the first five questions on the pre-/post-tests (the sixth question deliberately prompted students to connect classical and quantum mechanical reasoning). Each reasoning for each question can take on a value of -1, 0, or +1 depending on whether a particular reasoning (quantum or classical) was not good (-1), had some desirable and undesirable elements (0) or was  good (+1). Thus, the average change in the quality of reasoning on a particular question (defined as gain) can range from -2 to +2, indicating extremal negative change to extremal positive progress, respectively. More than 20\% of the sample was scored individually for inter-rater reliability  and the scores of the two raters agreed to better than 90\%.  
	

\begin{table}
\label{reasoningImprovement}
\caption{The gains (improvement from pre-test to post-test) on average correct quantum reasoning scores for 72 students, followed by the p-values (rounded to 2 digits).  The column ``Retention" shows that the gains are retained between the post-test and final exam on matched questions for 18 students.}
\begin{tabularx}{\columnwidth-4pt}{@{}| l | Y Y | Y Y |@{}}
\hline
Item 			& Gain		&p-value	&Retention	&p-value\\
1			&	+0.37	&0.01	&+0.06	&0.78\\
2			&	+0.41	&$<$0.01	&-0.06	&0.74\\
3			&	+0.24	&0.01	&+0.06	&0.68\\
4			&	+0.42	&$<$0.01	&{N/A}	&{N/A}\\
5			&	+0.34	&0.01	&+0.22	&0.43\\
\hline
\end{tabularx}
\end{table}

\subsection{Changes in Average Correct Reasoning}

	Table \ref{reasoningImprovement} displays changes in correct average quantum reasoning in student responses on each question from pre-test to post-test and whether those changes are statistically significant (all p-values $<$ 0.05).  Table \ref{reasoningImprovement} also includes a comparison of the average score using quantum reasoning on each question on the post-test (right after working on the QuILT) with quantum reasoning  on the four matched questions on the final exam several months later (Retention in column 4) and whether there are statistically significant differences between the two (p-values in the last column).  Table \ref{reasoningImprovement} does not include changes in average correct classical reasoning. However, a similar analysis for each question shows either no change or a very small positive gain in average correct classical reasoning from the pre-test to post-test and from the post-test to the final exam.
	
	Comparison of students' quantum reasoning in Table \ref{reasoningImprovement} shows significant gain in quantum mechanical reasoning from pre-test to post-test, and retention of these gains (change from the post-test to the final exam) more than two months later.   
The average scores on quantum reasoning on each matched question on the final exam compared to the post-test (right after the QuILT) suggest that knowledge was retained for more than two months on average.  
Since the average scores on the post-test and matched questions on the final exam are not statistically significantly different, we hypothesize that competing effects, e.g., the memory becoming rusty over time and consolidation of knowledge specific to quantum mechanics during the entire semester may both be playing a role in the final exam performance. 

\section{Summary}
	The Larmor precession QuILT focuses on helping students learn time-dependence of expectation values using a two state model. Data from several years show that student performance improved after working on the QuILT. Also, data on matched questions (same as pre-/posttest questions) administered to a subset of students on the final exam show that these gains are on average retained.  
\vspace*{-0.15in}
\begin{theacknowledgments}
	We thank the National Science Foundation for awards PHY-0968891 and PHY-1202909.
\end{theacknowledgments}
\vspace*{-0.15in}
\bibliographystyle{aipproc}   

\end{document}